\documentclass[a4paper,fleqn]{cas-dc}

\usepackage[sort,compress,numbers]{natbib}
\DeclareMathAlphabet{\mathcal}{OMS}{cmsy}{m}{n}
\def\tsc#1{\csdef{#1}{\textsc{\lowercase{#1}}\xspace}}
\tsc{WGM}
\tsc{QE}

\begin{document}



\title [mode = title]{Observation of topological phase with critical localization in a quasi-periodic lattice}

\author[1]{Teng Xiao}
\fnmark[1]
\author[1] {Dizhou Xie}
\fnmark[1]
\fntext[1]{These authors contributed equally to this work}
\author[1] {Zhaoli Dong}
\author[1]{ Tao Chen}
\affiliation[1]{Interdisciplinary Center of Quantum Information, State Key Laboratory of Modern  Optical Instrumentation, and Zhejiang Province Key Laboratory of Quantum Technology  and  Device of Physics Department, Zhejiang University, Hangzhou 310027, China}
\author[2,3]{ Wei Yi}
\cormark[1]
\affiliation[2]{CAS Key Laboratory of Quantum Information, University of Science and Technology of China, Hefei 230026, China}
\affiliation[3]{CAS Center For Excellence in Quantum Information and Quantum Physics, Hefei 230026, China}
\author[1,4]{Bo Yan}
\cormark[1]
\affiliation[4]{Collaborative Innovation Centre of Advanced Microstructures, Nanjing University, Nanjing, 210093, China}
\cortext[1]{Corresponding authors, Email address: wyiz@ustc.edu.cn (W.Y.) and yangbohang@zju.edu.cn (B.Y.)}



\begin{abstract}
Disorder and localization have dramatic influence on the topological properties of a quantum system. While strong disorder can close the band gap thus depriving topological materials of topological features, disorder may also induce topology from trivial band structures, wherein topological invariants are shared by completely localized states. Here we experimentally investigate a fundamentally distinct scenario where topology is identified in a critically localized regime, with eigenstates neither fully extended nor completely localized. Adopting the technique of momentum-lattice engineering for ultracold atoms, we implement a one-dimensional, generalized Aubry-Andr\'e model with both diagonal and off-diagonal quasi-periodic disorders in momentum space, and characterize its localization and topological properties through dynamic observables. We then demonstrate the impact of interactions on the critically localized topological state, as a first experimental endeavor toward the clarification of many-body critical phase, the critical analogue of the many-body localized state.
\end{abstract}



\maketitle


\section{Introduction}

Topologically protected edge modes in topological materials give rise to robust quantum charge transport against weak disorder~\cite{Hasan2010,Qi2011}. However, for sufficiently strong disorder, the band gap closes and all the eigenstates become localized under Anderson localization, rendering the system topologically trivial with vanishing transport~\cite{Anderson1958, Billy2008,Roati2008,Kondov2011}. Contrary to this well-accepted scenario, recent studies have shown that disorder can also induce topology in a trivial insulator, leading to a topological Anderson insulator where the global topology is carried by completely localized wave functions~\cite{Li2009, Guo2010}. Such intimate connections among topology, disorder, and localization stimulate intense research efforts in recent years~\cite{Jiang2009, Altland2014, Hughes2014,Liu2017,Agarwala2017, Chen2019, Tang2020}, which culminate in the experimental observation of disorder-induced topological Anderson insulators in cold atoms~\cite{Meier929} and photonics~\cite{Stutzer2018, Liu2020}.

Here we experimentally demonstrate an alternative scenario, where a topological phase exists in the critically localized regime.
Different from either the extended state (as in Bloch waves) or the fully localized state (as in Anderson localization), a critically localized state is characterized by critical features such as multifractal wave functions and singular continuous spectrum
~{\cite{Machida1986,Geisel1991,PhysRevB.50.11365,PhysRevB.55.12971,Takada2004,Mirlin2006,PhysRevB.91.014108,Wang2016,Bertrand2016,Cestari2016,Liu2016,PhysRevLett.126.080602,Wang2020}}.
Intuitively, such a regime can be viewed as an expansion of the extended-to-localized transition point in the standard Aubry-Andr\'e model, driven by an additional quasi-periodic disorder in the nearest-neighbor hopping rate.
Under many-body interactions, the non-interacting critical state becomes
an intriguing many-body critical phase~\cite{PhysRevLett.126.080602,Wang2020} that interpolates the extended, ergodic phase and the fully localized, nonergodic many-body localized phase~\cite{Schreiber842,Luschen2018, Rispoli2019,Abanin2019}.
While the critically localized state and its many-body incarnation have been theoretically investigated in several models with stochastic or quasi-periodic disorder~\cite{PhysRevB.50.11365,PhysRevB.55.12971,Takada2004,Mirlin2006,PhysRevB.91.014108,Bertrand2016,Cestari2016,Liu2016,Wang2016,PhysRevLett.126.080602,Wang2020},
their experimental demonstration has remained elusive, even for the non-interacting case.

Taking advantage of the flexible control available to ultracold atomic momentum-lattice engineering~\cite{Meier2016,An2018, An2018a, Xie2019, Xie2020}, we implement a generalized Aubry-Andr\'e model in one dimension, with quasi-periodic modulation in both the nearest-neighbor hopping rate and the on-site potential.
Using dynamic observables, we demonstrate indications for the existence of a critically localized topological state in the non-interacting regime, and characterize its competition with both the extended and fully localized phases.
We further explore the impact of interactions on the stability of the critical topological phase, paving the way toward a more systematic study of the many-body critical state.

\section{Results and discussion}

\begin{figure}[tbp]
\centering
\includegraphics[width= 0.45\textwidth]{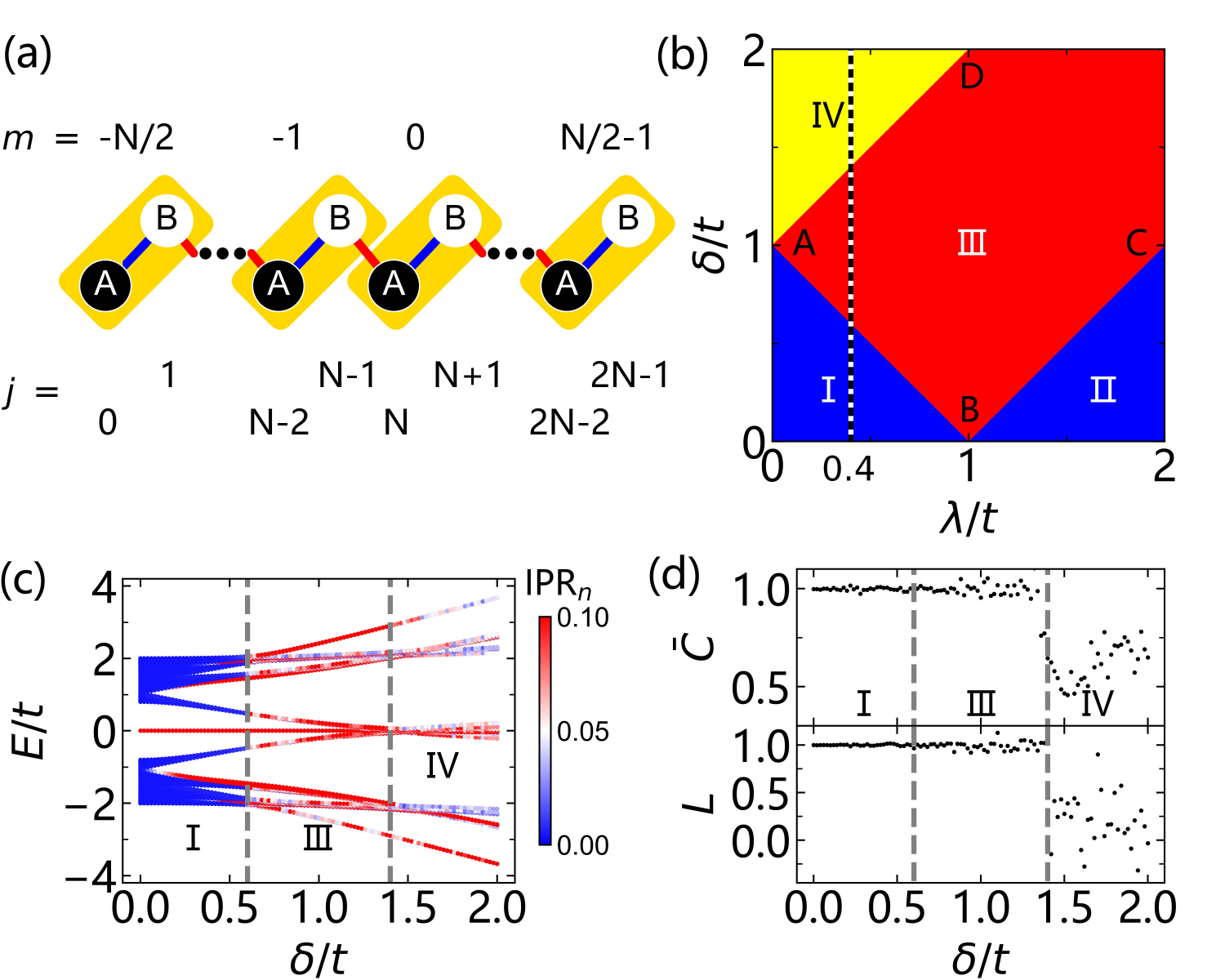}
\caption{\label{fig:model} (a) Schematic illustration of the generalized Aubry-Andr\'e model of Eq.~(\ref{eq:of}). The lattice sites $j$ are divided into unit cells (labeled by $m$) with sublattice sites $A$ and $B$.
(b) Theoretical phase diagram for $V=0$~\cite{Cestari2016, Liu2016}: regions I and II are the extended topological phase with Zak phase $\varphi_z=\pi$ (with our choice of unit cells); region III is the critically localized topological phase with $\varphi_z=\pi$; region IV is critically localized but with $\varphi_z=0$. Different regions are separated by linear boundaries: AB ($\delta /t= 1 - \lambda/t$), BC ($\delta/t = \lambda/t - 1$), and AD ($\delta/t = 1 + \lambda/t$).
(c) Eigenenergy spectrum with increasing off-diagonal disorder strength $\delta/t$, along the vertical dotted line in (b) with $\lambda/t=0.4$. Color code indicates the value of $\text{IPR}_n$, defined in the main text.
(d) Local topological marker and time-averaged mean chiral displacement (A-MCD) along the vertical dotted line in (b).
For numerical calculations in (c)(d), we take a lattice with $400$ sites ($N=200$) under an open boundary condition. Without loss of generality, we set $\varphi=0$ in (c), while taking the average over $50$ randomly generated $\varphi$ in the range of $[0,2\pi)$ in (d). The A-MCD is calculated by taking the time average over a time evolution initialized $|m=0,A\rangle$ and with a duration of $80~ \hbar/t$.
}
\end{figure}

{\bf Topology with critical localization.}
We consider a generalized Aubry-Andr\'e model with the Hamiltonian

\begin{align}
{\hat{H}}_{0} =& \sum_{j}\left\{ t + \lambda \cos[(j+1)\pi] + \delta \cos(2\pi{\beta}j + \varphi)\right\} ({\hat{c}}^{\dagger}_{j}{\hat{c}}_{j+1} \nonumber \\
&+\text{H.c.})+V\cos (2\pi\beta j) \hat{c}^\dag_j\hat{c}_j, \label{eq:of}
\end{align}
where $\hat{c}^\dag_j$ ($\hat{c}_j$) is the creation (annihilation) operator at site $j$.
As illustrated in Fig.~\ref{fig:model}(a), the model depicts a one-dimensional superlattice with intra- (inter-) cell hopping rates $t-\lambda$ ($t+\lambda$), respectively, and is further dressed by an on-site potential $V$, as well as an incommensurate hopping term with amplitude $\delta$ and phase $\varphi$.
We take $\beta = (\sqrt{5}-1)/2$ throughout the work, under which the incommensurate hopping (on-site potential) constitutes the off-diagonal (diagonal) disorder. We also take $\varphi=0$ throughout the work unless stated otherwise.

\begin{figure}[tbp]
\hspace{0.25\textwidth}
\includegraphics[width= 0.47\textwidth]{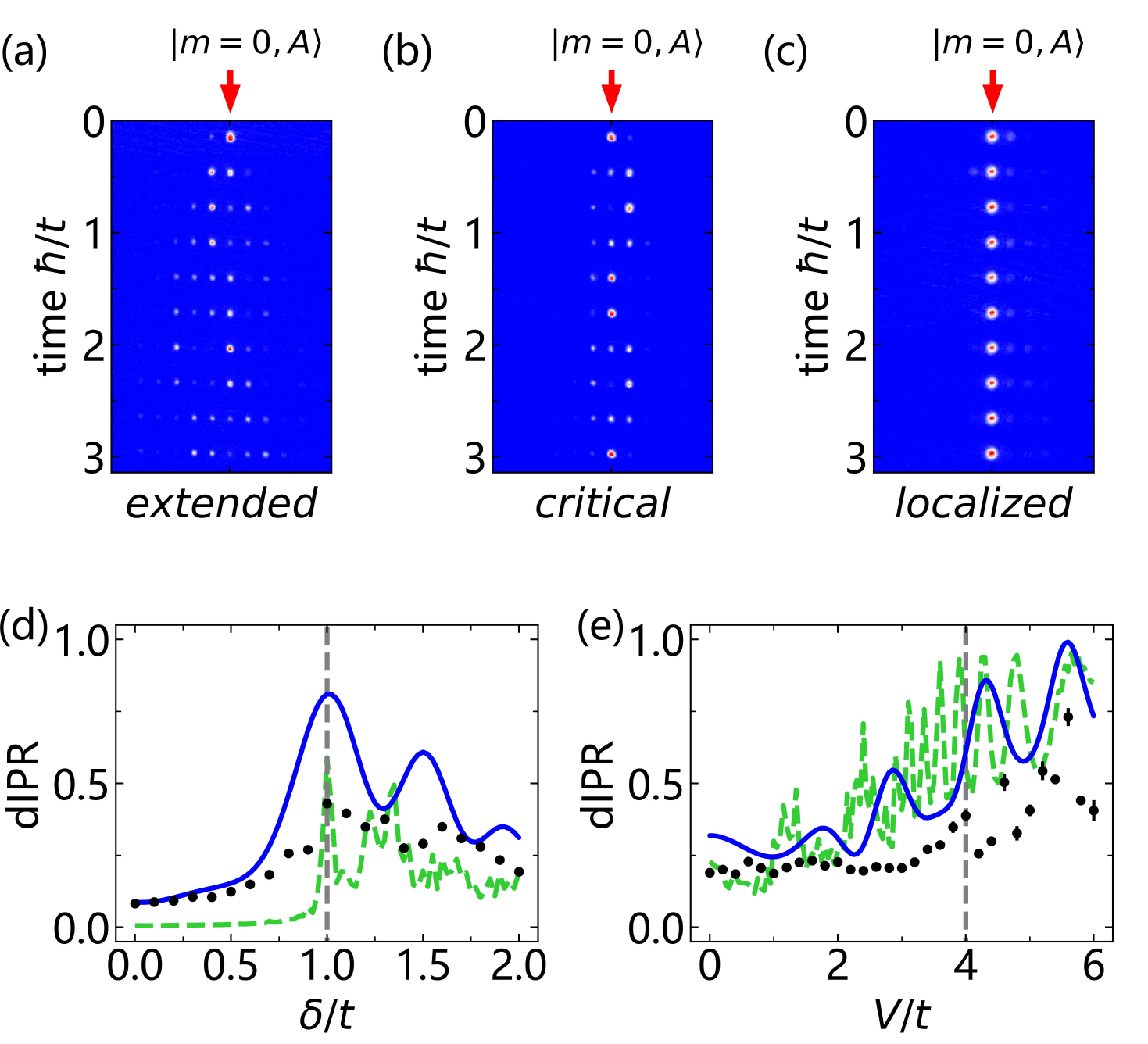}
\caption{\label{fig:IPR1}
(a)(b)(c) Measured time evolution of the atom-number distribution in momentum space in (a) the extended phase with $\lambda/t=0.4$ and $\delta=V=0$; (b) the critically localized state with $\lambda/t=0.4$, $\delta/t=1.2$, and $V=0$; (c) the fully localized phase with $\lambda=0$, $\delta/t=1.4$, and $V/t=5.6$.
(d) Extended to critical transition probed through dIPR with $\lambda/t=0$ and $V/t=0.2$. (e) Critical to fully localized transition with $\lambda=0$ and $\delta/t=2$. In (d)(e),
black dots are experimental data for an evolution time of $0.5$ ms along a momentum lattice with $16$ sites.
Blue line represents numerical results under the same parameters.
Green-dashed line is the numerical results with $400$ lattice sites, and with a longer evolution time of $80~\hbar/t$ ($\sim 12.7$ ms under experimental parameters). The vertical dashed lines are the theoretically predicted phase boundaries.
All numerical simulations in the main text are performed using Hamiltonian (\ref{eq:of})~\cite{supp}.
We fix $t/\hbar= 2\pi\times 1.0(1) $ kHz for all panels. Error bars show SE of mean.
}
\end{figure}

In the absence of disorder ($\delta=V=0$), Hamiltonian (\ref{eq:of}) is reduced to the standard Su-Schrieffer-Heeger model~\cite{Su1979}, where a topological phase transition occurs when $\lambda$ crosses zero. The topology of a Su-Schrieffer-Heeger model is characterized by the Zak phase $\varphi_z$, which takes quantized values of either $0$ or $\pi$, but is affected by the choice of unit cells.
Here we label the unit cells by requiring $\varphi_z=\pi$ for $\lambda>0$ in the absence of disorder, when edge states appear under open boundaries [see Fig.~~\ref{fig:model}(c)]. With the introduction of quasi-periodic disorder, a topological phase with critical localization emerges. In Fig.~\ref{fig:model}(b), we show the theoretical phase diagram in the presence of off-diagonal disorder ($\delta\neq 0$)~\cite{Cestari2016, Liu2016}, where a critically localized topological phase (region III, with $\varphi_z=\pi$) is surrounded by extended topological phases (regions I and II, with $\varphi_z=\pi$) and a critically localized phase with $\varphi_z=0$ (region IV). Note that an analytic mapping exists between the two extended regions I and II~\cite{Liu2016}. The phase diagram is further enriched when the diagonal disorder $V$ is turned on, as the critically localized state can become fully localized (Anderson localized)~\cite{Wang2020}. In the following, we first discuss the localization properties of the system.

For the $n$th eigenstate $|\Psi_n\rangle=\sum_j \psi_n(j)c^\dag_j|\text{vac}\rangle$ ($\psi_n(j)$ being the wave function), its localization property is captured by the inverse participation ratio, defined as $\text{IPR}_n=\sum_j |\psi_n(j)|^4$. Physically, $\text{IPR}_n$ is the inverse of the number of sites occupied by the $n$th eigenstate, with $\text{IPR}_n=0$ ($\text{IPR}_n=1$) for an extended (a fully localized) state. Such a behavior is rooted in the eigenstate's fractal dimension, defined in the thermodynamic limit as $\text{FD}=-\displaystyle{\lim_{N\rightarrow \infty}}\ln (\text{IPR}_n)/\ln (2N)$~\cite{RevModPhys.80.1355}, where $N$ is the number of unit cells. The fractal dimension is $1$ ($0$) for an extended (fully localized) state, consistent with the scaling behavior of the inverse participation ratio. More important, as a manifestation of its multifractal nature, a critically localized state has a fractal dimension in between $0$ and $1$, ensuring that its $\text{IPR}_n$ is also in the same range, for a sufficiently large system~\cite{Cestari2016,Liu2016,PhysRevLett.126.080602,Wang2020}. Indeed, across the localization transition boundary [AB in Fig.~\ref{fig:model}(b)], $\text{IPR}_n$ typically changes from zero in region I, to finite values in between $0$ and $1$ in regions III and IV [see Fig.~\ref{fig:model}(c)].
Note that, apart from a handful of edge states, our model does not possess mobility edges~\cite{supp}.

To characterize topological properties, we adopt the local topological marker~\cite{Hughes2014,Meier929}, defined through
\begin{align}
L=\frac{1}{2}\text{Tr}'(\hat{Q}\hat{\Gamma} \hat{m} \hat{Q}),
\end{align}
where $\text{Tr}'$ is the trace over a single unit cell deep in the bulk, $\hat{m}$ is the unit-cell position operator, $\hat{Q}=\hat{P}_+-\hat{P}_-$, $\hat{\Gamma}=\hat{\Gamma}_A-\hat{\Gamma}_B$, where $\hat{P}_{+}$ ($\hat{P}_-$) is the projection operator onto the band with positive (negative) eigenenergies, and $\hat{\Gamma}_A$ ($\hat{\Gamma}_B$) projects onto the sublattice site $A$ ($B$) [see Fig.~\ref{fig:model}(a)].
The topological marker can be dynamically probed through the mean chiral displacement (MCD)~\cite{Meier929,Cardano2017, Meier2016a}, defined as
\begin{align}
C=2\langle \hat{\Gamma} \hat{m}\rangle,
\end{align}
which is the expectation value of the chiral displacement operator $\hat{\Gamma} \hat{m}$ with respect to the time-evolved state. Experimentally, we use the time-averaged mean chiral displacement (A-MCD) $\bar{C}$ to characterize different topological phases, which converges faster in dynamics~\cite{Xie2020}. Here A-MCD is defined as $\bar{C}=\frac{1}{N}\sum_n C(n\Delta\tau)$, where the total evolution time is evenly discretized into $N$ sections, and we average over the measured MCD at each discrete time $n\Delta\tau$ ($n=1,...,N$). For all our measured $\bar{C}$, we take $N=10$ for an evolution time of $0.5$ ms. For numerical simulations, we take $N=50$ for the same duration of evolution.
As shown in Fig.~\ref{fig:model}(d), the numerically evaluated local topological marker and A-MCD are both close to the quantized value of $1$ in regions I and III, revealing their topological nature.
A topological phase transition occurs on the boundary of regions III and IV, as $L$ and $\bar{C}$ are no longer quantized in region IV. Interestingly, neither $L$ nor $\bar{C}$ vanishes in region IV, due to the apparent divergence of the localization length therein~\cite{supp}.

{\bf Experimental implementation.}
We experimentally implement Hamiltonian (\ref{eq:of}) by engineering a one-dimensional momentum lattice of $16$ sites ($N=8$) in a $^{87}$Rb Bose-Einstein condensate~\cite{Xie2019, Gou2020,supp}. As discrete momentum states are coupled by a series of frequency-modulated Bragg-laser pairs, both diagonal and off-diagonal disorder are imposed by adjusting the Bragg-coupling parameters between adjacent sites.
For the time evolution, atoms are initialized in the lattice site $|m=0,A\rangle$. We then pulse on the Bragg lasers and evolve the condensate on the momentum lattice. For detection, the dipole trap and Bragg laser beams are switched off, and atoms are allowed to expand freely for $20$ms.
As atoms in different momentum states are naturally separated in space, we then take an absorption image, from which all dynamic quantities are extracted. The interaction is typically weak in our experiment, though the relative ratio of the interaction and kinetic energy can be tuned by adjusting the atom density. We adopt such a scheme later to characterize the impact of interaction.

\begin{figure}
\hspace{0.25\textwidth}
\includegraphics[width= 0.47\textwidth]{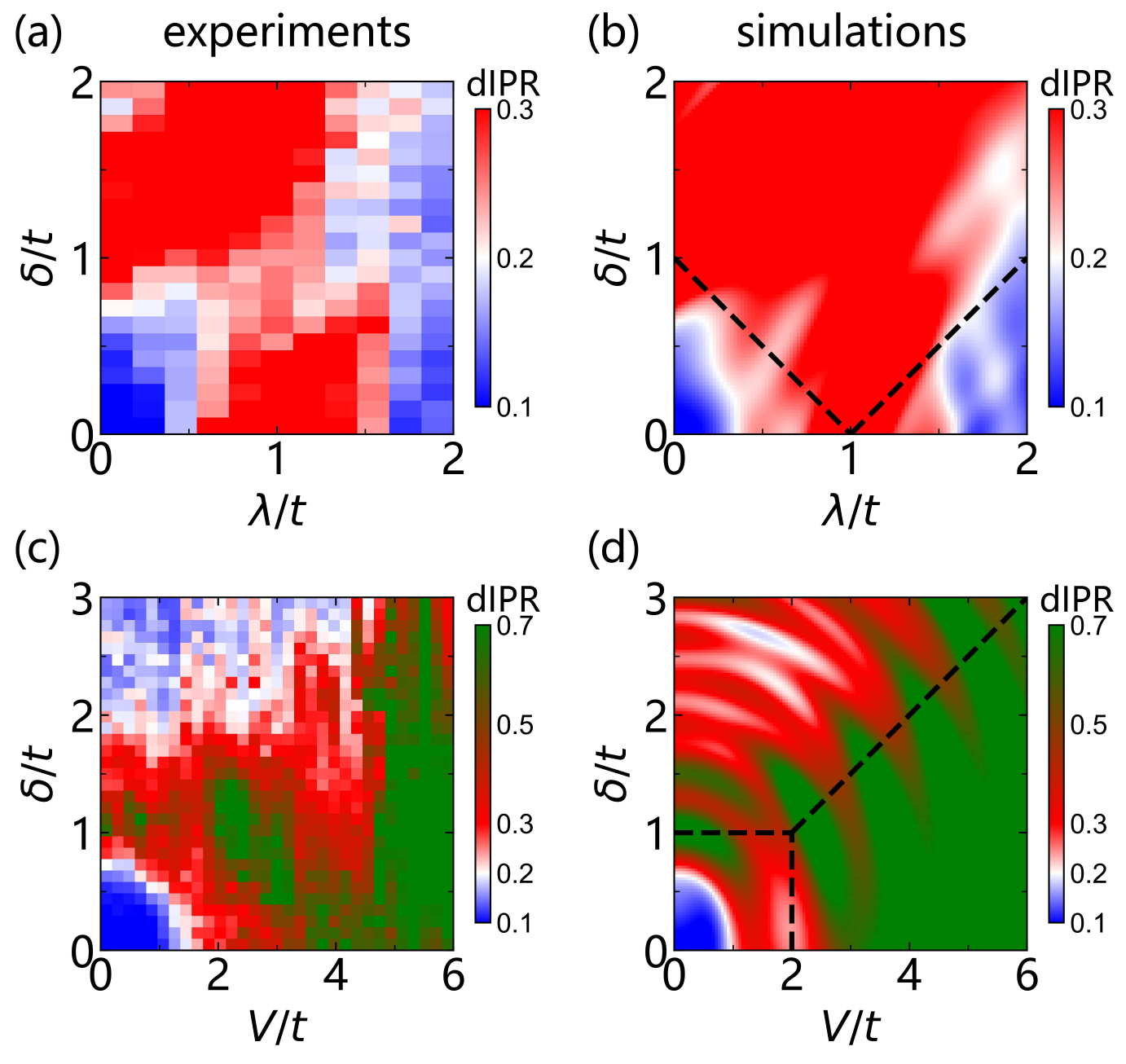}
\caption{\label{fig:IPR2}
(a) Color contour of measured dIPR on the $\delta$--$\lambda$ plane, with $V=0$.
(b) Numerically calculated dIPR under the same parameters as (a).
(c) Color contour of measured dIPR on the $\delta$--$V$ plane, with $\lambda=0$.
(d) Numerically calculated dIPR under the same parameters as (c). Here we fix $t/\hbar = 2\pi\times 1.0(1) $ kHz along a lattice with $16$ sites, and with an evolution time of $0.5$ ms. Theoretically predicted phase boundaries are shown in (b)(d) using black dashed lines~\cite{PhysRevB.50.11365,PhysRevB.91.014108,PhysRevLett.126.080602}.
}
\end{figure}

{\bf Localization properties.}
Signatures of localized eigen wave functions are qualitatively observed in the dynamic propagation of the condensate.
In the absence of disorder ($\delta=V=0$), all eigen wave functions are extended, apart from edge states which have negligible overlap with the initial state. This leads to the ballistic expansion observed in Fig.~\ref{fig:IPR1}(a). By contrast, for sufficiently large $\delta$, all the eigen wave functions are critically localized, giving rise to the observed quasi-localization, where the condensate, no longer exponentially localized, spills over to adjacent unit cells near the initial site [see Fig.~\ref{fig:IPR1}(b)].
This differs markedly from a fully localized state under sufficiently large diagonal disorder $V$, where only the initial site is visibly occupied under an exponential localization [see Fig.~\ref{fig:IPR1}(c)].

More quantitatively, we construct the dynamic inverse participation ratio (dIPR) from the time-evolved state $|\Psi(t)\rangle=\sum_j \psi(j,t) c_j^\dag|\text{vac}\rangle$ as
\begin{align}
\text{dIPR}=\sum_j |\psi(j,t)|^4,
\end{align}
where $\psi(j,t)$ is the time-evolved wave function on site $j$.
A vanishing (finite) dIPR at long times indicates the extended (critical or local) nature of eigenstates that have support on the initially occupied lattice site $|m=0,A\rangle$. Due the absence of mobility edge in our model~\cite{PhysRevLett.126.080602,supp}, dIPR should reflect the localization property of the eigenenergy spectrum, and provide a practical probe to the aforementioned $\text{IPR}_n$ and fractal dimension
which are only rigorously defined in the thermodynamic limit. For instance, an extended to critical transition can be identified
near the theoretically predicted transition point $\delta/t=1$, in the long-time numerical simulation of dIPR [green-dashed curve in Fig.~\ref{fig:IPR1}(d)]. The experimentally measured dIPR (dots) also show a peak near the predicted transition point.
For sufficiently large $\delta$, the observed dIPR appears to oscillate around a finite value well below unity, indicating the critically localized nature of the state.
Such a behavior is to be contrasted with that under a large diagonal disorder $V$.
As shown in Fig.~\ref{fig:IPR1}(e), the measured dIPR starts from a small but finite value indicative of the critically localized state, grows appreciably as $V$ increases, and approaches unity for sufficiently large $V$, where the system is fully localized. Here the theoretically predicted transition point is $V/t=4$~\cite{PhysRevB.50.11365,PhysRevB.91.014108,PhysRevLett.126.080602}.

\begin{figure}[tbp]
\includegraphics[width= 0.47\textwidth]{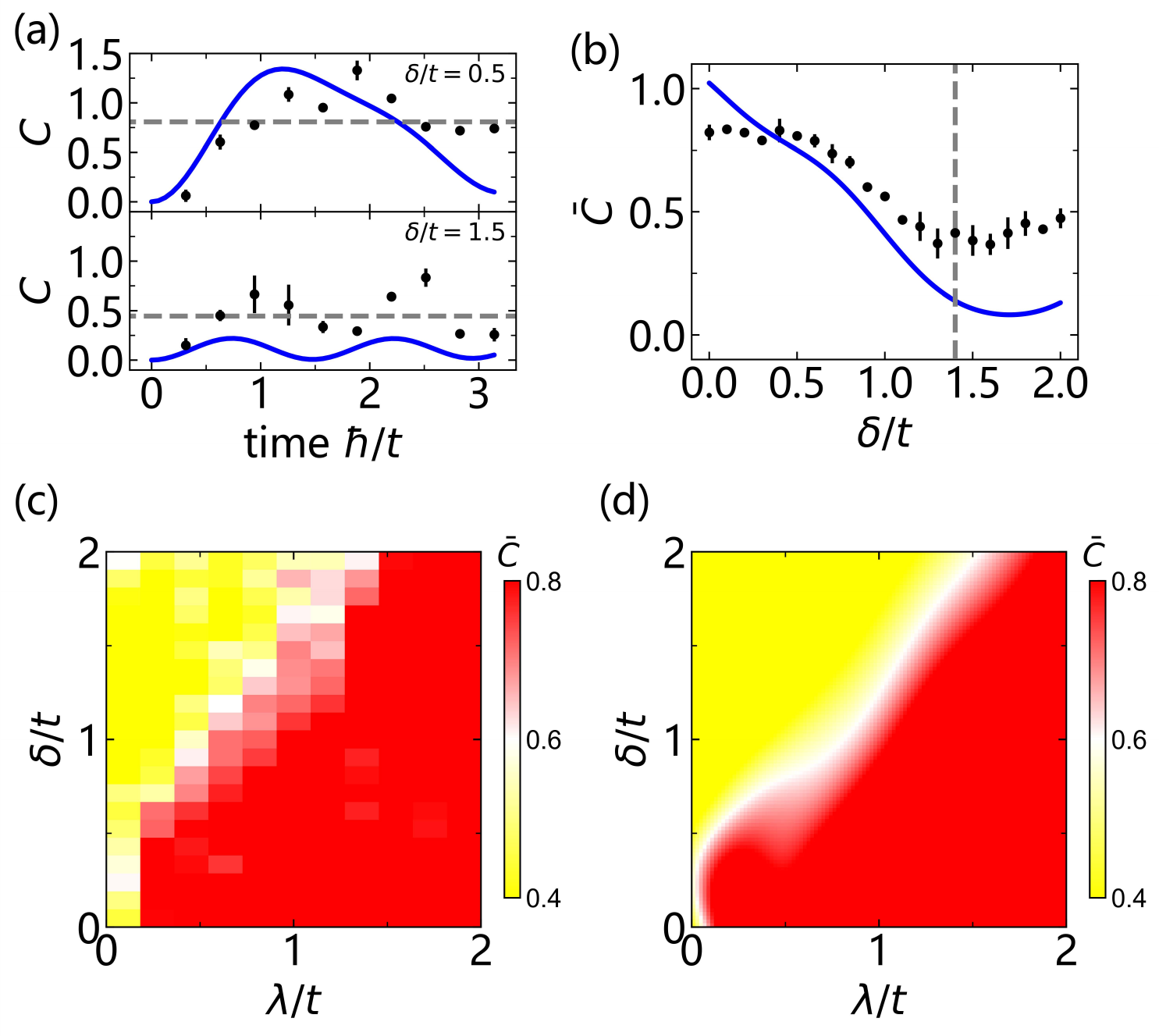}
\caption{\label{fig:mcd} (a) Time evolution of MCD in region I (upper panel with $\delta/t=0.5$), and region IV (lower panel with $\delta/t=1.5$). The gray lines are the averages of experimental data.
(b) A-MCD $\bar{C}$ as a function of $\delta/t$. The gray vertical line indicates the theoretical phase transition point $\delta/t=1.4$.
In (a)(b), black dots are experimental data (error bars show SE of mean), blue lines are from numerical simulations, and we fix $\lambda/t=0.4$.
(c) Color contour of experimentally measured A-MCD.
(d) Numerically calculated A-MCD.
Other parameters are the same as those in Fig.~\ref{fig:IPR2}.
}
\end{figure}

\begin{figure}[tbp]
\includegraphics[width= 0.47\textwidth]{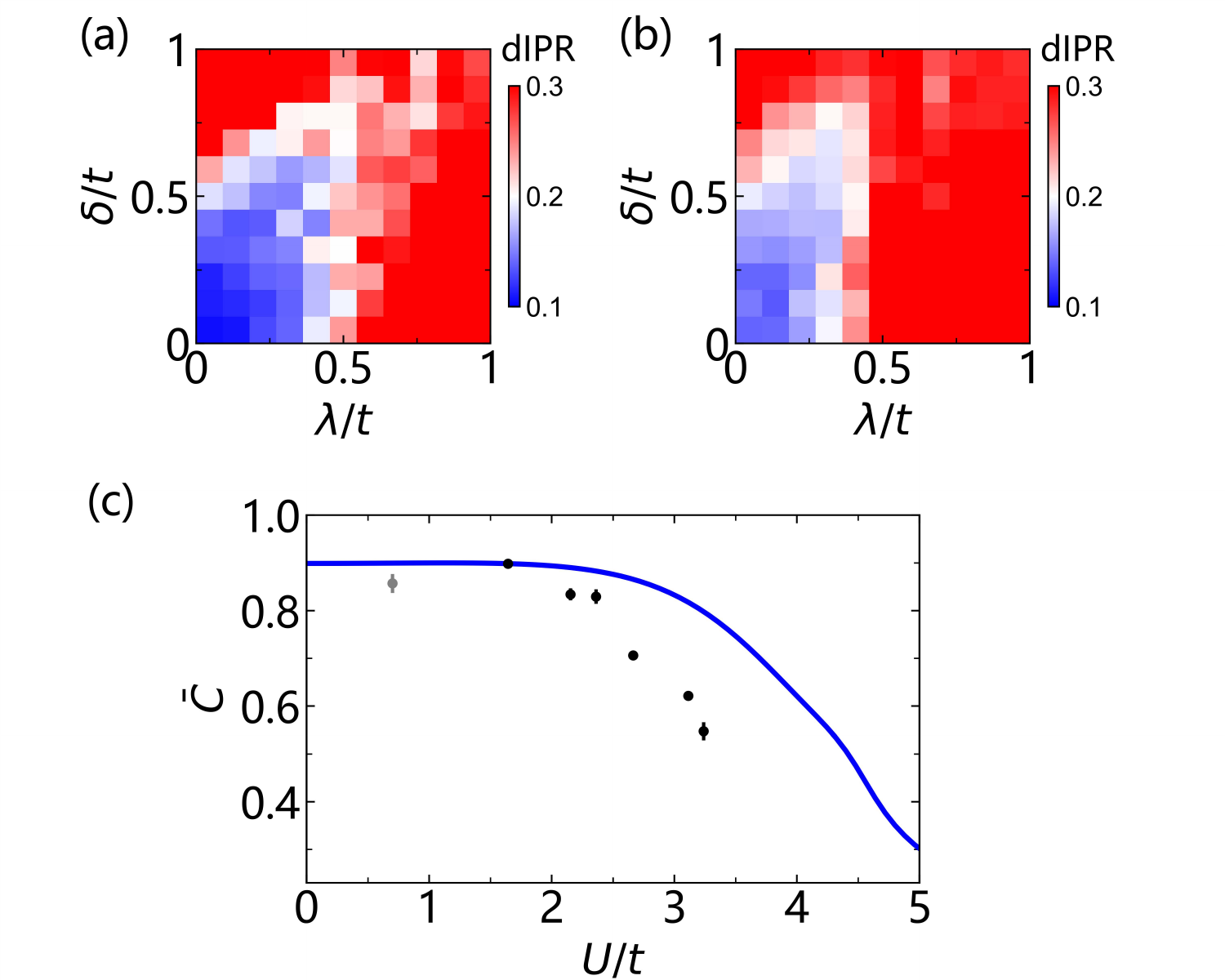}
\caption{\label{fig:U}
(a)(b) Experimentally measured dIPR under different interaction energy $U$, with (a) $U/t=1.6$ and (b) $U/t =3.2$. (c) A-MCD as a function of $U/t$ for $\lambda/t =0.2$ and $\delta/t =0.2$. Black dots are experimental data and the blue line is the numerical result under the same parameters. Both experimental and numerical results are performed on a lattice of $16$ sites, with $t/\hbar = 2\pi\times 0.50(5) $ kHz and a total evolution time of $1$ ms.
The gray dot in (c) represents the corresponding experimental data in Fig.~\ref{fig:mcd}, with $U/t=0.7$. Error bars show SE of mean.}
\end{figure}

Boundaries between different phases are easier to identify in Fig.~\ref{fig:IPR2}, where we show color-contoured phase diagrams using the measured dIPR.
In Fig.~\ref{fig:IPR2}(a), we plot the experimental data on the $\delta$--$\lambda$ plane with $V=0$, where the critical to extended phase transitions are consistent with the theoretical prediction in Fig.~\ref{fig:model}(b). In Fig.~\ref{fig:IPR2}(c), the measured phase diagram is shown on the $\delta$--$V$ plane, with $\lambda=0$. All three phases: the extended (blue), the critical (red white), and the fully localized (green) phases show up in Fig.~\ref{fig:IPR2}(c), with their phase boundaries in qualitative agreement with theoretical predictions [black dashed in Fig.~\ref{fig:IPR2}(d)]~\cite{PhysRevB.50.11365,PhysRevB.91.014108,PhysRevLett.126.080602}. Qualitatively similar phase diagrams can also be constructed by probing other dynamic observables~\cite{supp}.

The accuracy of the measured phase diagram is limited by the finite evolution time, which is in turn set by the decoherence time. In our experiment, the decoherence is caused by imperfections in the system, such as off-resonant Bragg couplings, the inhomogeneity caused by the dipole trap, and inhomogeneous interactions between atoms~\cite{Xie2020, Gou2020}. Simulations taking some of these imperfections into account would provide a better fit to experimental data~\cite{supp}.

{\bf Topological phase diagram.}
We now focus on the topological properties of the system under $V=0$, where a critically localized topological phase exists.
In Fig.~\ref{fig:mcd}(a), we show the time evolution of MCD under different parameters. The observed MCD oscillates around unity in region I, whereas in region IV, it fluctuates around $0.5$. A topological phase transition can be identified through the A-MCD [see Fig.~\ref{fig:mcd}(b)], close to the theoretically predicted topological transition point $\delta/t=1.4$.
Similar to the case with dIPR, we map out the topological phase diagram in Fig.~\ref{fig:mcd}(c)(d). Our measured A-MCD clearly indicates a topological phase transition at sufficiently large $\delta$ [Fig.~\ref{fig:mcd}(c)], with the boundary of the transition consistent with both the numerical results [Fig.~\ref{fig:mcd}(d)] and theoretical prediction [Fig.~\ref{fig:model}(b)].

{\bf Impact of interaction.}
The inter-atomic collisions in our system can be effectively modeled, on the mean-field level, as attractive on-site interactions governed by the Hamiltonian~\cite{Xie2019,An2018,supp}
\begin{equation}
{\hat{H}}_{\text{int}} = U \sum_{j}( 2 - \langle \hat{n}_{j}\rangle) {\hat{n}}_{j}. \label{eq:u}
\end{equation}
Here, the number operator ${\hat{n}}_{j} = \frac{1}{N_t}{\hat{c}}^{\dagger}_{j}{\hat{c}}_{j}$, $N_t$ is the total number of atoms, and
$\langle \hat{n}_j\rangle$ is the normalized atomic population on site $j$, with $\sum_j\langle \hat{n}_j\rangle=1$.
We tune the ratio between the interaction and kinetic energies $U/t$ by adjusting the density of atoms~\cite{supp}.

Figures \ref{fig:U}(a)(b) contrast experimentally measured localization phase diagrams on the $\delta$--$\lambda$ plane, for different values of $U/t$ and $V=0$. In previous experiments, $U/t=0.7$, and the impact of interaction is not apparent~\cite{supp}.
With a larger $U/t$, the stability region of the extended state becomes smaller, while the region of critical localization increases. This is consistent with previous studies of interaction effects in momentum lattices, where interactions facilitate localization~\cite{Xie2020}. 
On the other hand, in the topologically non-trivial regime [see Fig.~\ref{fig:U}(c)], the measured A-MCD ($\bar{C}$) remains quantized up to $U/t\sim 2$, indicating the robustness of the topological phase against small interactions.
Under stronger interactions, $\bar{C}$ appreciably deviates from unity, suggesting an interaction-induced topological phase transition.

While we mainly focus on region III in the phase diagram, the entire region IV features a non-quantized MCD and a divergent localization length~\cite{supp}, in marked contrast to Anderson topological insulators where localization length only diverges on the topological phase boundary~\cite{Li2009, Guo2010,Jiang2009, Altland2014, Hughes2014,Liu2017,Agarwala2017, Chen2019, Tang2020}.
This is an interesting example where the interplay of topology and localization can lead to surprising results.
In future studies, it would also be desirable to incorporate Feshbach resonance~\cite{Marte2002,Syassen2006,Thalhammer2005} to offer more flexible control over the strength and sign of interactions.

\section*{Conflict of interest}
The authors declare that they have no conflict of interest.

\section*{Acknowledgments}

We acknowledge the support from the National Key Research and Development Program of China under Grant Nos. 2018YFA0307200,  2016YFA0301700 and 2017YFA0304100, the National Natural Science Foundation of China under Grant Nos. 12074337 and 11974331, Natural Science Foundation of Zhejiang province under Grant Nos. LR21A040002 and LZ18A040001, Zhejiang Province Plan for Science and technology No. 2020C01019 and the Fundamental Research Funds for the Central Universities.

\bibliographystyle{naturemag}
\bibliography{GAAHdraft_1022}

\end{document}


\title{Observation of topological phase with critical localization in a quasi-periodic lattice}%

\author{ Teng Xiao}
\thanks{These authors contributed equally to this work}
\author{Dizhou Xie}
\thanks{These authors contributed equally to this work}
\author{Zhaoli Dong}
\author{Tao Chen}
\affiliation{%
Interdisciplinary Center of Quantum Information, State Key Laboratory of Modern Optical Instrumentation, Zhejiang Province Key Laboratory of Quantum Technology and Device, Department of Physics, Zhejiang University, Hangzhou 310027, China
}%
\author{Wei Yi}
\email{wyiz@ustc.edu.cn}
\affiliation{CAS Key Laboratory of Quantum Information, University of Science and Technology of China, Hefei 230026, China}
\affiliation{CAS Center For Excellence in Quantum Information and Quantum Physics, Hefei 230026, China}
\author{Bo Yan}
\email{yanbohang@zju.edu.cn}
\affiliation{%
Interdisciplinary Center of Quantum Information, State Key Laboratory of Modern Optical Instrumentation, Zhejiang Province Key Laboratory of Quantum Technology and Device, Department of Physics, Zhejiang University, Hangzhou 310027, China
}%
\affiliation{%
Collaborative Innovation Centre of Advanced Microstructures, Nanjing University, Nanjing, China, 210093
}%
\affiliation{ Key Laboratory of Quantum Optics, Chinese Academy of Sciences, Shanghai, China, 200800
}

\appendix
\pagebreak
\widetext
\begin{center}
\textbf{\large Supplemental Material for ``Observation of topological phase with critical localization in a quasi-periodic lattice''}
\end{center}
\setcounter{equation}{0}
\setcounter{figure}{0}
\setcounter{table}{0}
\makeatletter
\renewcommand{\theequation}{S\arabic{equation}}
\renewcommand{\thefigure}{S\arabic{figure}}
\renewcommand{\bibnumfmt}[1]{[S#1]}
\renewcommand{\citenumfont}[1]{S#1}

\setcounter{equation}{0}
\setcounter{figure}{0}
\setcounter{table}{0}
\makeatletter
\renewcommand{\theequation}{S\arabic{equation}}
\renewcommand{\thefigure}{S\arabic{figure}}
\renewcommand{\citenumfont}[1]{#1}

In this Supplemental Material, we provide details for the experimental setup, derivation of the full and effective Hamiltonians, features of the edge states, theoretical analysis of topological number and localization length, details on the calculation of mean chiral displacement, additional experimental data supporting the critically localized phase, and interaction effects.

\section{\label{app:subsec}Experimental implementation of the generalized Aubry-Andr\'e model}

We experimentally implement Hamiltonian (1) in the main text by engineering a one-dimensional momentum lattice of $16$ sites ($N=8$) in a $^{87}$Rb Bose-Einstein condensate (BEC) of $\sim 10^5$ atoms in a dipole trap with trapping frequencies $2\pi\times (50,160, 170)$Hz. Before switching on the Bragg lasers that couple different momentum states, we adiabatically decompress the trapped gas along the lattice direction, such that the atoms are only weakly trapped along the direction of the lattice, but are strongly trapped in other directions. This also serves to minimize the wave-function overlap of different momentum states, so that dynamics in the momentum space is coherent, and we may regard discrete momentum-lattice sites as distinguishable quantum states~\cite{Meier2016,An2018, An2018a}.

\begin{figure}[h]
\includegraphics[width= 0.55\textwidth]{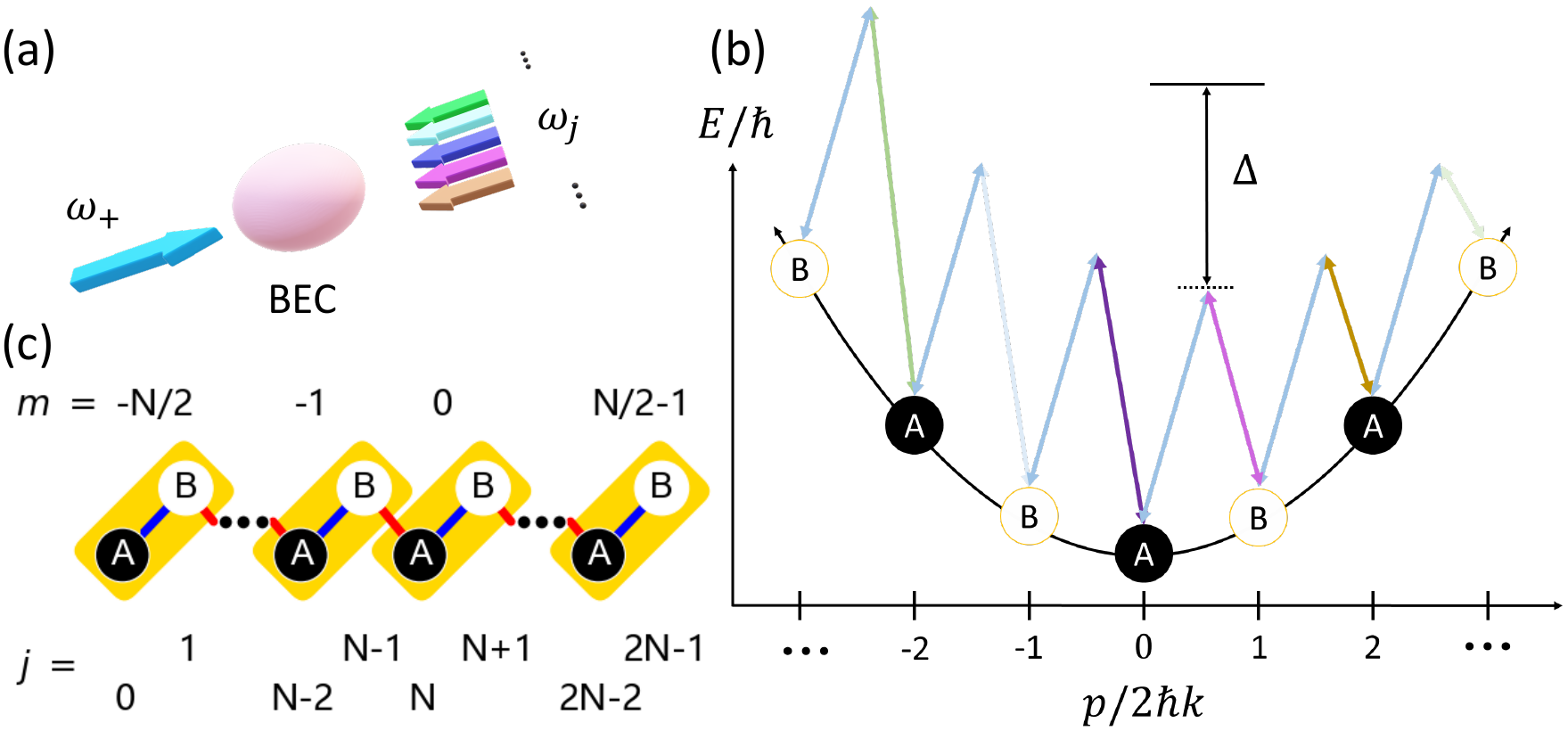}
\caption{\label{fig:S1} Generating the one-dimensional momentum lattice. (a) Bragg-coupling scheme. (b) Pairs of counter-propagating Bragg lasers with fine-detuned frequencies ($\omega_+$ and $\omega_j$) couple adjacent momentum lattice sites. (c) A schematic illustration of the resulting one-dimensional momentum lattice.
}
\end{figure}

Discrete momentum states along the weakly trapped direction are then coupled by pairs of Bragg lasers.
By engineering the frequencies of the multi-component Bragg lasers, a one-dimensional momentum lattice with adjacent sites separated by $2\hbar k$ is created, whose nearest-neighbor hopping rates can be tailored by individually tuning the coupling strengths of the Bragg processes. Here $k$ is the wave number of the $1064$nm laser.
Both diagonal and off-diagonal disorder are individually tuned by adjusting the Bragg-coupling parameters between adjacent momentum-lattice sites. For instance, the off-diagonal disorder is implemented by modulating the amplitude of the Bragg beam with frequency $\omega_j$ with the function $[1+\lambda/t\cos(j+1)\pi+\delta/t\cos(2\pi\beta j+ \varphi)]$ (see main text for definition of variables). Here the frequency $\omega_j$ is tuned to the two-photon resonance between the corresponding momentum states [see Fig.~\ref{fig:S1}(b)], via an acousto-optic modulator. Likewise, the diagonal disorder is implemented by setting the two-photon detunings of the Bragg processes.

While for most of our experiments, we adopt a hopping amplitude $t/\hbar= 2\pi\times 1.0(1)$kHz, we fix $\varphi=0$ for all experiments.
Without further tuning, the inter-atomic interaction energy $U/\hbar\approx 2\pi\times 0.7$kHz, whose impact is small as $U<t$. Here $U=g\rho$, where $g=4\pi\hbar^2 a/M$, $\rho$ is the average density of the initial BEC evaluated with the Thomas-Fermi radius, $a=103a_0$ is the $s$-wave scattering length ($a_0$ being the Bohr radius), and $M$ is the atomic mass.

For the time evolution, atoms in the condensate are naturally initialized in the lattice site $|m=0,A\rangle$.
We then switch on the Bragg lasers and let the condensate evolve in the momentum lattice, after we adiabatically decompress the trapped gas along the lattice direction. At the end of the time evolution, we turn off the dipole trap and Bragg lasers, and take an absorption image of the atoms after $20$ms time of flight in free space. This allows us to probe the atomic density distribution along the momentum lattice right before the time-of-flight, from which all dynamic quantities are extracted.

Finally, to study the interaction effects (as in Fig.~5 of the main text), we vary the ratio between the interaction and kinetic energies $U/t$ by adjusting the total density of BEC. In order to achieve a relatively large tuning range, we fix a smaller hopping rate $t= 2\pi\times 0.50(5)$kHz, and hold BEC in the dipole trap with larger trapping frequencies $ 2\pi\times(70, 240, 250)$Hz. The atomic density is varied by holding the atoms in the dipole trap for different durations (few to tens of seconds). As single-photon scattering removes atoms from the trap, we end up with condensates with different densities.
In this way, we minimize other effects, and achieve a wider tuning range of $U/t$ ($0.7\sim 3.2$).
For future studies, the Feshbach resonance technique can be employed to control both the sign and magnitude of the interaction rate $U$.


\section{Derivation of the full and effective Hamiltonians}

To analyze the impact of off-resonant Bragg couplings, we first write down the full and effective Hamiltonians. Following the derivation in Ref.~\cite{Gou2020},  the full Hamiltonian, containing both resonant and off-resonant couplings, can be written in the interaction picture as
\begin{align}
H_\text{f} =\sum\limits_n \sum_{j} \frac{\hbar\tilde\Omega_j}{2}e^{i8(n-j)E_r\tau/\hbar}e^{-i\delta_j\tau}e^{-i\phi_j}|n+1\rangle\langle n| + \text{H.c.},
\label{eq:suppfullH}
\end{align}
where the recoil energy $E_r=\hbar^2 k^2/2M$, $\tilde{\Omega}_j$ and $\delta_j$ are respectively the two-photon Rabi frequency and two-photon detuning of the $j$th Bragg process, $\phi_j$ is the phase of the $j$th Bragg coupling, $|n\rangle$ is the momentum-lattice site with momentum $2n\hbar k$, and $\tau$ is time.

When we neglect the off-resonant terms by setting $n=j$, we get the effective Hamiltonian
\begin{equation}
H_\text{eff} =\sum\limits_n \frac{\hbar\tilde\Omega_n}{2}e^{-i\delta_n\tau}e^{-i\phi_n}|n+1\rangle\langle n| + \text{H.c.}.
\end{equation}
 
We now go back to the Schr\"odinger's picture with the unitary transformation $e^{iH_0t}$, where $H_0=V_n |n\rangle\langle n|$.
Here, $V_n$ is defined through $\delta_n=(V_n-V_{n+1})/\hbar$. We then define $t_n=\frac{\hbar\tilde\Omega_n}{2}=t_j$, and $\phi_n=0$, so that
\begin{equation}
H_\text{eff} =\sum\limits_n t_n (|n+1\rangle\langle n| + \text{H.c.}) +V_n |n\rangle\langle n| .
\label{eq:suppeffH}
\end{equation}

In our experiment, we tune the parameters of the Bragg processes, such that $t_n=t + \lambda\cos(n+1)\pi + \delta\cos(2\pi\beta{n}+\varphi)$, and $V_n=V\cos(2\pi\beta{n})$, which recovers Hamiltonian (1) in the main text. Specifically, the off-diagonal (characterized by $t_n$) and the diagonal (characterized by $V_n$) quasiperiodic disorders are implemented, respectively, by tuning the effective Rabi frequencies ($\tilde{\Omega}_j$) and two-photon detunings ($\delta_j$) of the Bragg couplings between any two adjacent sites along the lattice.
For numerical simulations in the main text, we use the effective Hamiltonian Eq.~(\ref{eq:suppeffH}). In the following sections of this Supplemental Materials, simulations with both the
full (\ref{eq:suppfullH}) and effective (\ref{eq:suppeffH}) Hamiltonians are presented to show their difference, mainly due to off-resonant coupling terms neglected in Eq.~(\ref{eq:suppeffH}).

\begin{figure}[h]
\includegraphics[width= 0.65\textwidth]{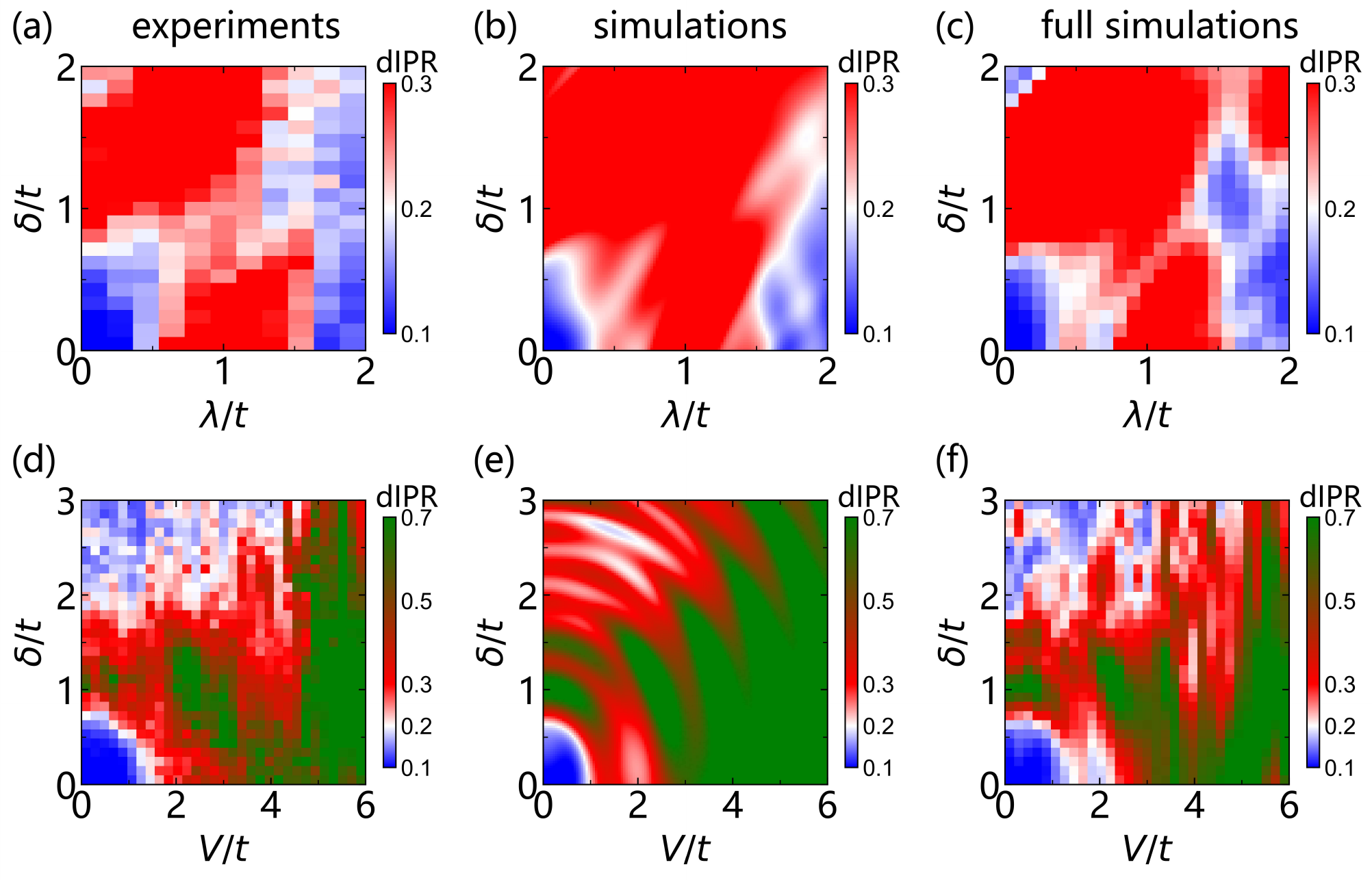}
\caption{\label{fig:S_full1}
Localization phase diagram.
(a) Color contour of measured dIPR on the $\delta$--$\lambda$ plane, with $V=0$.
(b) Numerically calculated dIPR under the same parameters as (a), using the effective Hamiltonian.
(c) Numerically calculated dIPR with full Hamiltonian under the same parameters as (a).
(d) Color contour of measured dIPR on the $\delta$--$V$ plane, with $\lambda=0$.
(e) Numerically calculated dIPR under the same parameters as (d), using the effective Hamiltonian.
(f) Numerically calculated dIPR with full Hamiltonian under the same parameters as (d).
Here we fix $t/\hbar = 2\pi\times 1.0(1) $kHz along a lattice with $16$ sites, and with an evolution time of $0.5$ms.
}
\end{figure}

\section{Impact of off-resonant couplings}

As a concrete example, we show in Fig.~\ref{fig:S_full1}, the comparison between the experimentally measured dynamic inverse participation ratio (dIPR) [Fig.~\ref{fig:S_full1}(a)(d), same as Fig.~3(a)(c) of the main text], numerical simulations with the effective Hamiltonian (\ref{eq:suppeffH}) [Fig.~\ref{fig:S_full1}(b)(e), same as Fig.~3(b)(d) of the main text], and simulations with the full Hamiltonian (\ref{eq:suppfullH}) [Fig.~\ref{fig:S_full1}(c)(f)]. Simulations with the full Hamiltonian provides a more satisfactory fit for the experimental data. Nevertheless, all cases are consistent with the theoretically predicted phase diagrams (see black dashed lines in Fig.~3 of the main text).

\section{\label{app:subsec}Zero-energy and isolated edged states}

\begin{figure}[tbp]
\includegraphics[width= 0.55\textwidth]{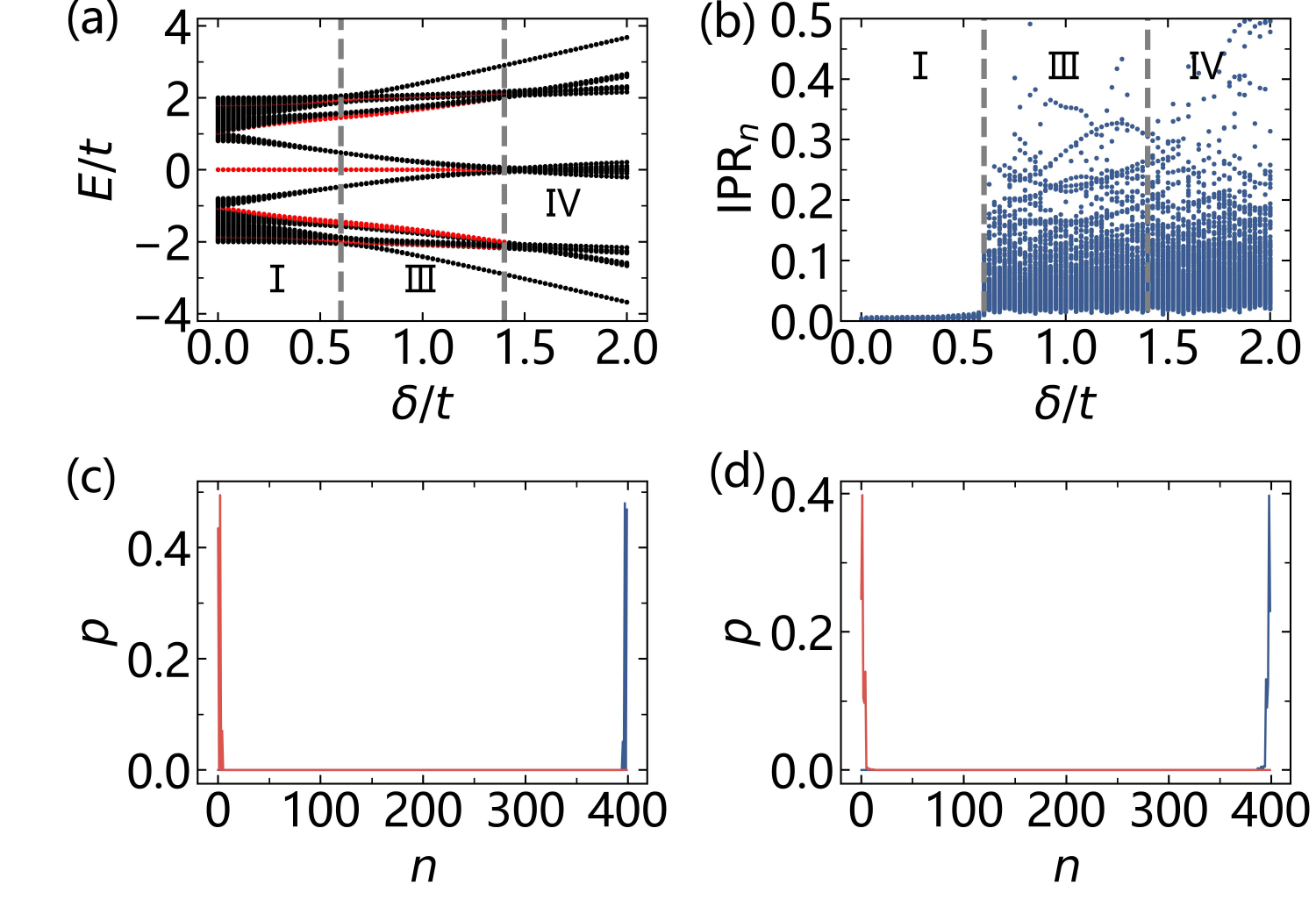}
\caption{\label{fig:S2} Eigenenergy spectrum and edge states of Hamiltonian (1) of the main text. (a) Eigenenergy spectrum for $\lambda/t=0.4$, with edge states (both zero-energy edge states and other isolated edge states) marked in red. (b) $\text{IPR}_n$ for all bulk eigenstates (edge states excluded) with increasing disorder strength $\delta/t$ for $\lambda/t=0.4$. (c) Density distribution ($p$) of zero-energy edge states with $\lambda/t=0.4$ and $\delta/t=0.5$. (d) Density distribution ($p$) of a typical pair of isolated edge states with $\lambda/t=0.4$ and $\delta/t=0.5$. For all panels, we take $\varphi=0$.
}
\end{figure}

In Fig.~\ref{fig:S2}(a), we show the complete eigenenergy spectrum under an open boundary condition, with $\lambda/t =0.4$.
Both the zero-energy and isolated edge states are shown in red, with their density distributions plotted in Fig.~\ref{fig:S2}(c) (zero-energy), and Figs.~\ref{fig:S2}(d) (isolated), respectively. The zero-energy edge states are topological, with their appearance/disappearance consistent the topological phase diagram shown in Fig.~1(b) of the main text. From the $\text{IPR}_n$ calculation in Fig.~\ref{fig:S2}(b) [as well as in Fig.~1(c) of the main text], our model does not possess mobility edges, since the extended to critical localization transition occurs at the same point for all eigenstates (excluding the edge states). Further, beyond the
extended to critical transition at $\delta/t=0.6$, the calculated $\text{IPR}_n$ for all eigenstates become finite [see Fig.~\ref{fig:S2}(b)], but still appreciably smaller than unity, confirming their critically localized nature.


\section{\label{app:subsec}Mean chiral displacement}

\begin{figure}[tbp]
\includegraphics[width= 0.705\textwidth]{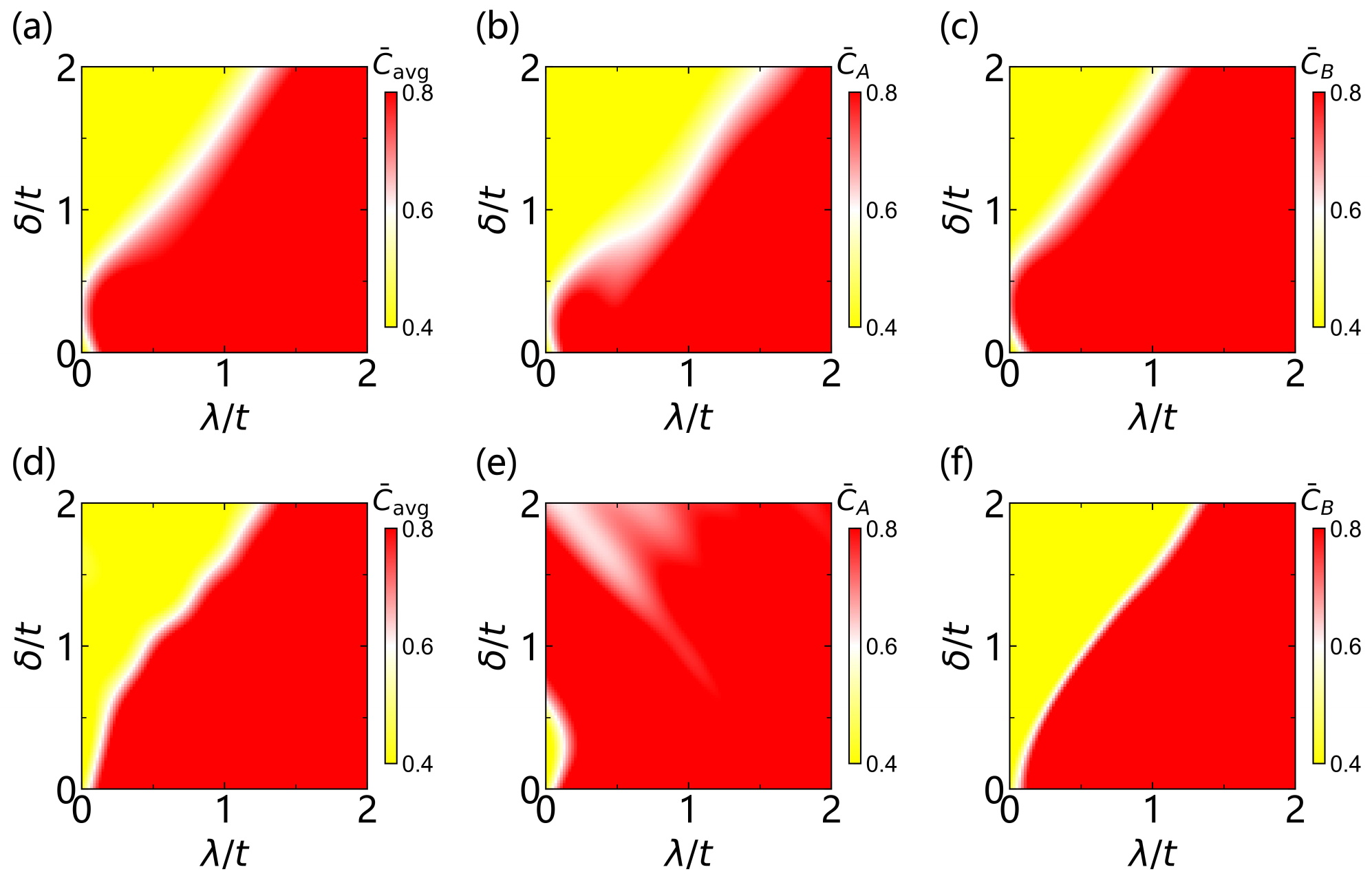}
\caption{\label{fig:S3} Numerical simulation of different definitions of A-MCDs, $\bar{C}_{\text{avg}}$, $\bar C_A$, and $\bar C_B$. (a)(b)(c) Comparison of A-MCDs on a lattice of $16$ sites ($N=8$). (d)(e)(f) Comparison of A-MCDs on a lattice of $28$ sites ($N=14$).
For all calculations, we take an overall evolution time of $0.5$ms with $t/\hbar = 2\pi\times1 $kHz and $\varphi=0$, consistent with experimental parameters. The effective Hamiltonian is used for all simulations.
}
\end{figure}

For our measurements of time-averaged mean chiral displacement (A-MCD) in the main text, the atoms are initialized on the site $|m=0,A\rangle$. More generally, A-MCD is sensitive to the initially populated sublattice site. To see this point, we define
\begin{align}
\bar{C}_{\sigma}(\tau) =\frac{2}{N}\sum_{n=1}^N \langle m=0,\sigma |e^{i\hat{H}_0n\Delta\tau}(\hat\Gamma\hat{m})e^{-i\hat{H}_0n\Delta\tau}|m=0,\sigma\rangle,
\end{align}
where $\sigma\in\{A,B\}$ indicates the initial sublattice site, $\tau$ stands for the total evolution time, $\Delta\tau=\tau/N$ is the length of each discretized time sector, and we take $N=50$ for numerical simulations.
In Fig.~\ref{fig:S3}, we show the numerically calculated A-MCDs $\bar{C}_{\sigma}$, as well as $\bar{C}_{\text{avg}}=(\bar{C}_A+\bar{C}_B)/2$ on the $\delta$--$\lambda$ plane under different parameters. While under experimental parameters, $\bar{C}_A$, $\bar{C}_B$ and $\bar{C}_{\text{avg}}$ are qualitatively the same and are consistent with the theoretical phase diagram [see Fig.~\ref{fig:S3}(a)(b)(c)]; for other choices of parameters, $\bar{C}_A$ and $\bar{C}_B$ can be quite different [see Fig.~\ref{fig:S3}(d)(e)(f)]. It is thus preferable to use $\bar{C}_{\text{avg}}$ to characterize the A-MCD in general. Alternatively, we find that, by averaging over randomly chosen $\varphi$, $\bar{C}_A$ and $\bar{C}_B$ would converge to the same values [see Fig.~1(d) in main text]. In particular, the peculiar contour of $\bar{C}_A$ in Fig.~\ref{fig:S3}(e) would average out to be similar to that of $\bar{C}_{\text{avg}}$, over randomly chosen $\varphi$.

\section{Dynamic phase diagram in terms of other observables}

To justify our choice of dynamic observable, and to confirm the existence of critically localized state,
we also adopt other dynamic quantities to probe the phase diagram, with consistent results.

\begin{figure}[htbp]
\includegraphics[width= 0.47\textwidth]{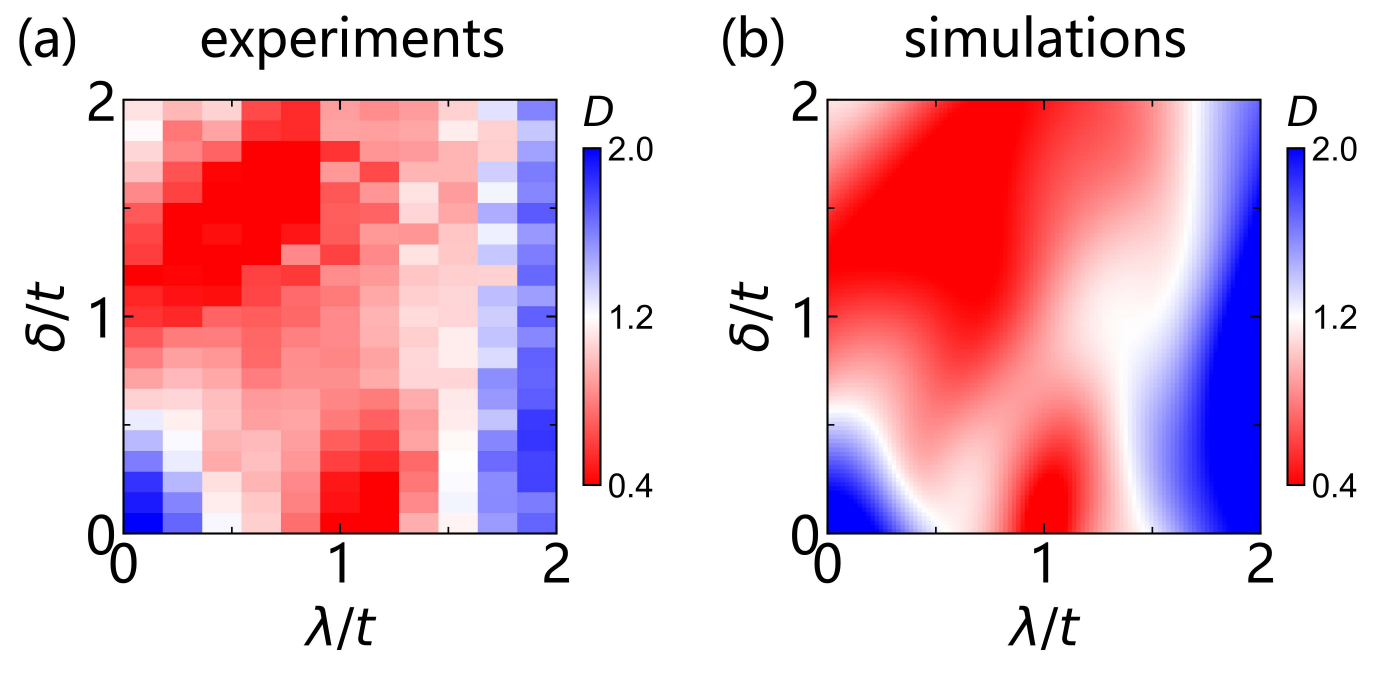}
\caption{\label{fig:S_D}
Localization phase diagram in terms of the spread of wave packet $D$.
(a) Color contour of measured $D$ on the $\delta$--$\lambda$ plane.
(b) Numerically calculated $D$ under the same parameters as (a), using the effective Hamiltonian.
Here we fix $t/\hbar = 2\pi\times 1.0(1) $kHz along a lattice with $16$ sites, and with an evolution time of $0.5$ms.
}
\end{figure}

The first dynamic observable, is the spread of wave packet, defined as
\begin{align}
D=\bigg[\sum_{m,\sigma=A,B}m^2 n_{m,\sigma}\bigg]^{1/2},
\end{align}
where $n_{m,\sigma}$ is the fraction of atom population on sublattice site $\sigma$ of unit cell $m$. We show our experimental data and numerical results in Fig.~\ref{fig:S_D}(a)(b), respectively, which qualitatively agree with the phase diagram in Fig.~3(a)(b) of the main text.

\begin{figure}[htbp]
\includegraphics[width= 0.705\textwidth]{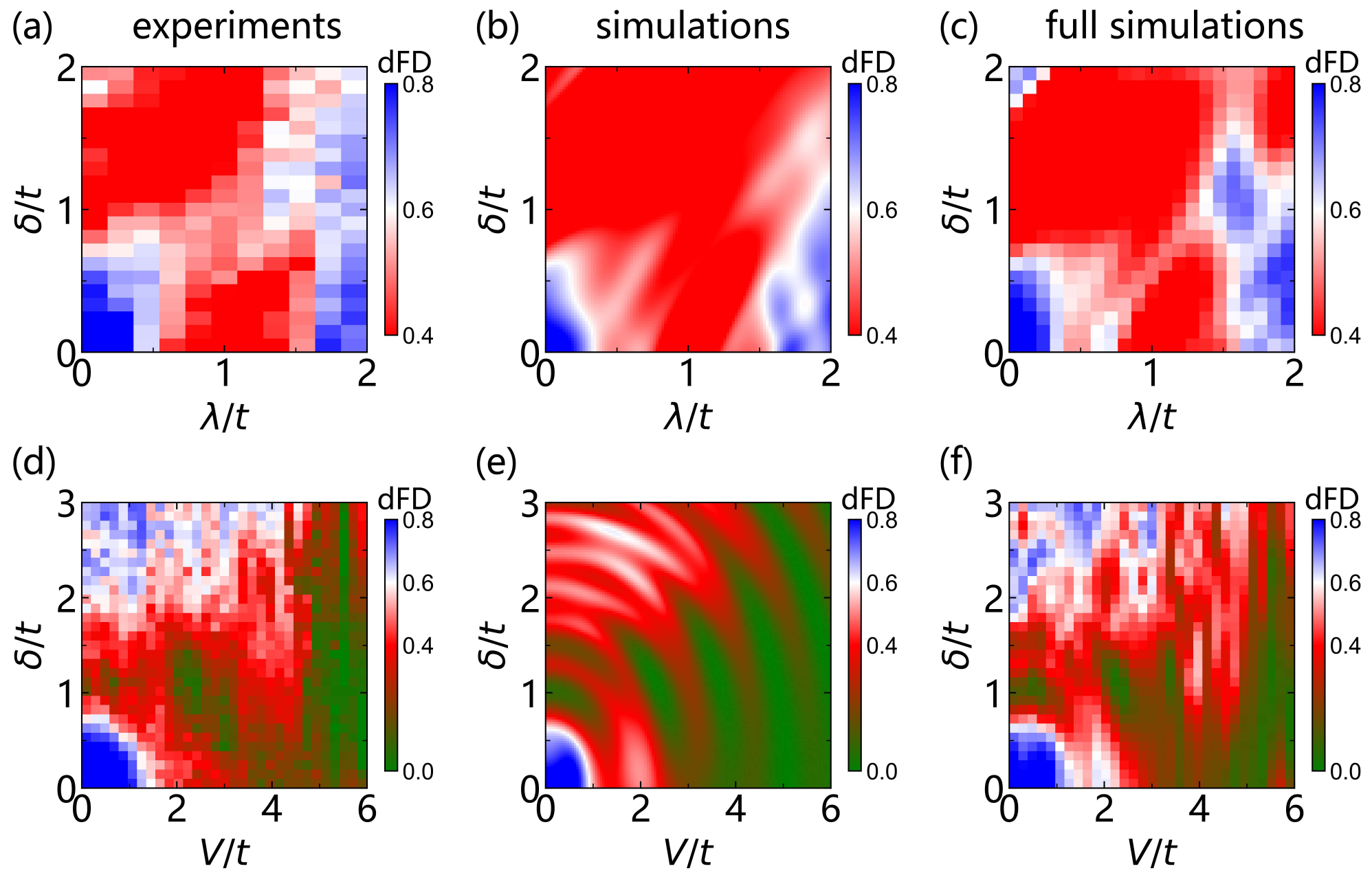}
\caption{\label{fig:S_dFD}
Localization phase diagram in terms of dFD.
(a) Color contour of measured dFD on the $\delta$--$\lambda$ plane with $V=0$.
(b) Numerically calculated dFD under the same parameters as (a), using the effective Hamiltonian.
(c) Numerically calculated dFD with full Hamiltonian under the same parameters as (a).
(d) Color contour of measured dFD on the $\delta$--$V$ plane with $\lambda=0$.
(e) Numerically calculated dFD under the same parameters as (d), using the effective Hamiltonian.
(f) Numerically calculated dFD with full Hamiltonian under the same parameters as (d).
Here we fix $t/\hbar = 2\pi\times 1.0(1) $kHz along a lattice with $16(N=8)$ sites, and with an evolution time of $0.5$ms.
}
\end{figure}

Alternatively, we define the dynamic fractal dimension
\begin{align}
\text{dFD}=-\frac{\ln \text{dIPR}}{\ln (2N)},
\end{align}
which mirrors the relation between the fractal dimension and the inverse participation ratio. Ideally, for a sufficiently large system, the $\text{dFD}$ should correctly reflect the scaling relation between the fractal dimension and the inverse participation ratio. While our experimental system is not very large in size ($N=8$), the measured dynamic phase diagrams, as shown in Fig.~\ref{fig:S_dFD}(a)(d), is again in agreement with the one measured using the $\text{dIPR}$, both consistent with theoretical predictions. Note that we also show the numerically simulated dFD using the full Hamiltonian Eq.(\ref{eq:suppfullH}) [see Fig.~\ref{fig:S_dFD}(c)(f)], which provide a better fit to the experimental data in Fig.~\ref{fig:S_dFD}(a)(d), compared to simulations using the effective Hamiltonian [see Fig.~\ref{fig:S_dFD}(b)(e)]. Our series of experiment thus strongly supports the presence of a critically localized phase.

\section{Dynamic phase diagram from time evolutions with longer duration}

\begin{figure}[tbp]
\includegraphics[width= 0.705\textwidth]{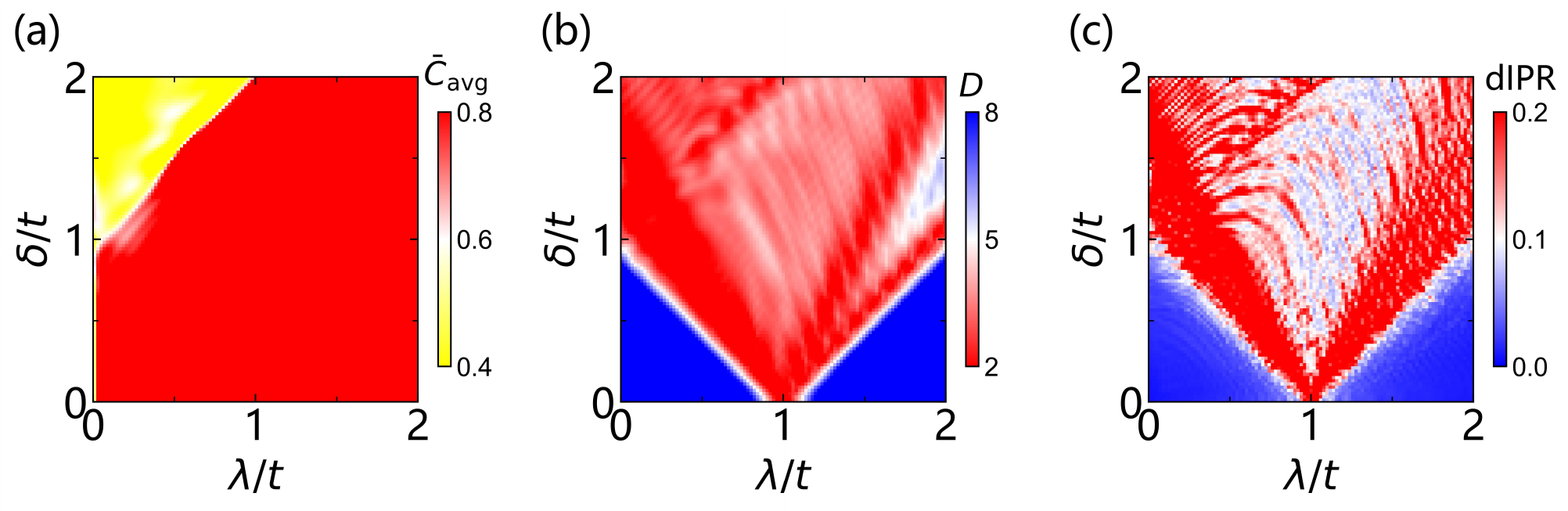}
\caption{\label{fig:S4} Numerical simulation of (a) $\bar{C}_{\text{avg}}$, (b) $D$, and (c) $\text{dIPR}$ in the $\delta$--$\lambda$ plane, for an evolution with the duration $60\hbar/t$ on a lattice of $320$ ($N=160$) sites. For the calculations, we fix the hopping rate as $t/\hbar = 2\pi\times1 $kHz and take $\varphi=0$. The effective Hamiltonian is used for simulations.
}
\end{figure}

A main factor that limits the accuracy of the dynamically determined phase diagrams is the finite evolution time achievable in our experiment, which is constrained by the decoherence time of the Bragg-coupled BEC. To explicitly demonstrate this, we numerically simulate a time evolution with a much longer duration, driven by the Hamiltonian (1) (the effective Hamiltonian), and plot the various dynamic quantities in Fig.~\ref{fig:S4}. Apparently, given a sufficiently long evolution time, both the topological [Fig.~\ref{fig:S4}(a)] and localization phase diagrams [Fig.~\ref{fig:S4}(b)(c)] are well-captured by dynamic quantities such as A-MCD, spread of wave packets $D$, and dIPR, as all phase diagrams are consistent with the theoretically predicted one in Fig.~1(b) of the main text.

Similarly, in Fig.~\ref{fig:S_critical}, we show the numerically evaluated dIPR on the $\delta$-$V$ plane with different values of $\lambda$, under the effective Hamiltonian. For the $\lambda=0$ case, the color-contour phase diagram is consistent with the theoretical prediction in Refs.~\cite{PhysRevB.50.11365,PhysRevB.91.014108,PhysRevLett.126.080602}. For the $\lambda/t=0.4$ case, both the fully localized and extended phases are visibly compressed.


\section{Localization phase diagram with a finite $\lambda$}

\begin{figure}[htbp]
\includegraphics[width= 0.705\textwidth]{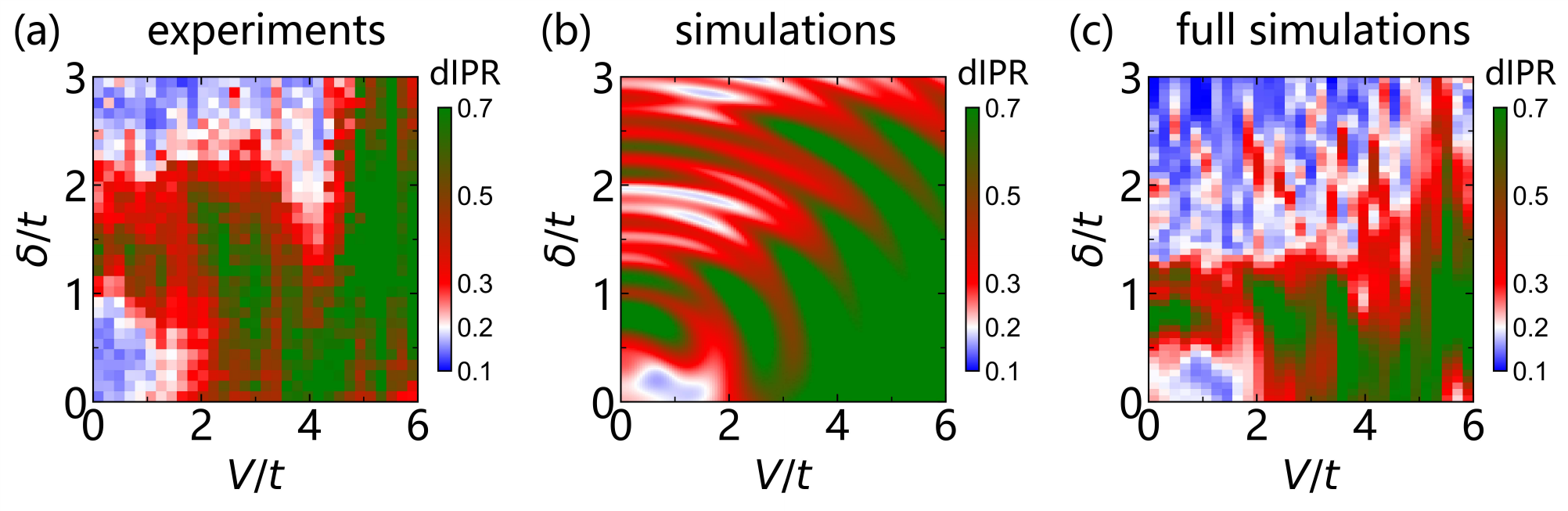}
\caption{\label{fig:S_full2}
Localization phase diagram.
(a) Color contour of measured dIPR on the $\delta$--$V$ plane, with $\lambda=0.4$.
(b) Numerically calculated dIPR under the same parameters as (a), using the effective Hamiltonian.
(c) Numerically calculated dIPR with full Hamiltonian under the same parameters as (a).
Here we fix $t/\hbar = 2\pi\times 1.0(1) $kHz along a lattice with $16$ sites, and with an evolution time of $0.5$ms.
}
\end{figure}

\begin{figure}[h]
\includegraphics[width= 0.47\textwidth]{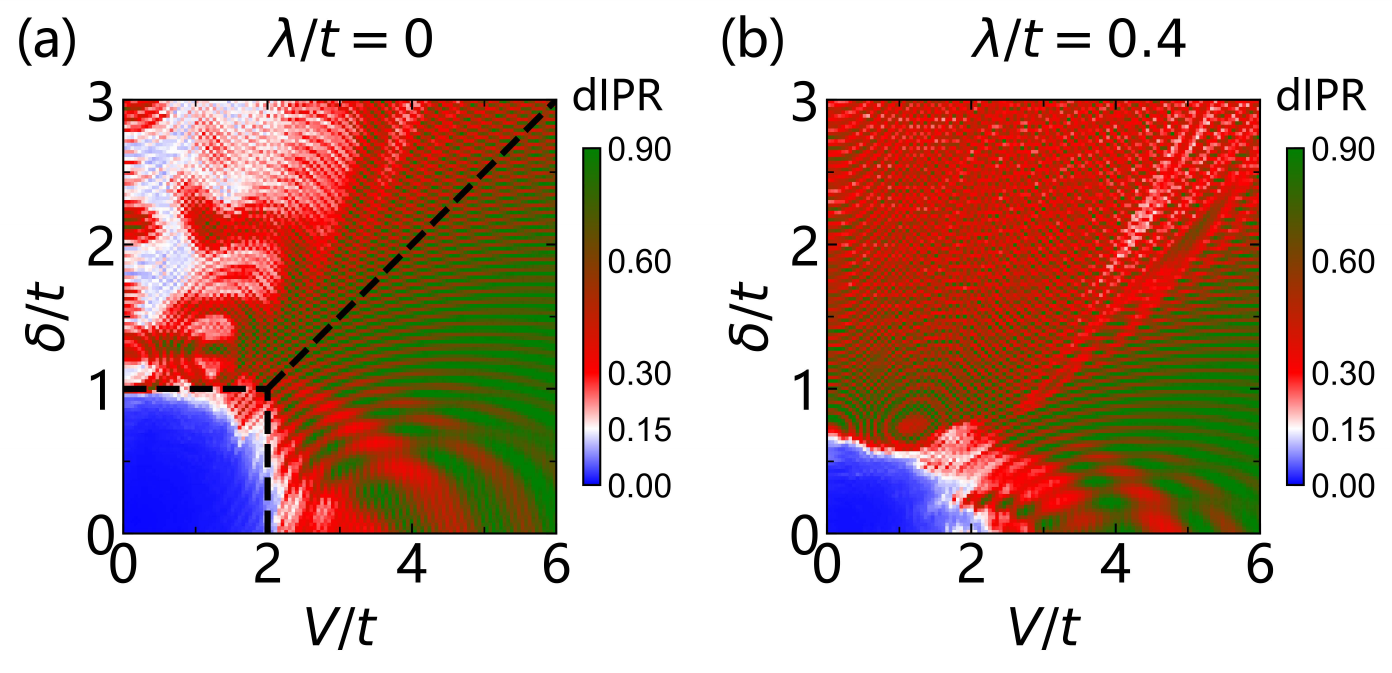}
\caption{\label{fig:S_critical}
Numerical simulation of dIPR in the $\delta$--$V$ plane, for an evolution with the duration of $80 \hbar/t$ on a lattice of $400$ ($N=200$) sites.
Here we fix $t/\hbar = 2\pi\times 1 $kHz, with the parameters (a) $\lambda=0$ and (b) $\lambda/t=0.4$. Theoretically predicted phase boundaries are shown in (a) using black dashed lines~\cite{PhysRevB.50.11365,PhysRevB.91.014108,PhysRevLett.126.080602}. Here we use the effective Hamiltonian for the simulations.
}
\end{figure}

In Fig.~3(c) of the main text, we measure the localization phase diagram on the $\delta$--$V$ plane with $\lambda=0$, which are consistent with previous theoretical predictions~\cite{PhysRevB.50.11365,PhysRevB.91.014108,PhysRevLett.126.080602}. While previous theoretical study only focused on the case of $\lambda=0$, we also experimentally probe the phase diagram for a finite $\lambda$. This is shown in Fig.~\ref{fig:S_full2}.
Compare to the $\lambda=0$ case, the measured phase boundaries are only slightly shifted. See Fig.~\ref{fig:S_critical} for numerical simulations with a longer evolution time and a larger lattice, where the difference is more apparent.

\section{\label{app:subsec}Topological number and localization length of the generalized Aubry-Andr\'e model}

We now discuss the topological phase boundary on the $\delta$--$\lambda$ plane with $V=0$.
While both the local topological marker (LTM) and A-MCD are able to characterize the topological phase boundary, in the critically localized, topologically trivial region (region IV), LTM and A-MCD seem to spread around a half-integer value of $0.5$ rather than around zero. Such a phenomenon is related to the apparent divergence of the localization length as well as the closing of the band gap in the entire region IV.

According to Refs.~\cite{arXiv.2008.06493,PhysRevB.83.155429}, the topological number $\nu$ of our model can be written as
\begin{equation}
\nu=\frac{1}{2}(1-\nu'),
\end{equation}
where
\begin{align}
\nu'&=\text{sign}\bigg\{\prod_j{\big[1-\lambda+\delta \cos(2\pi{\beta}\times 2j)\big]}^2
- \prod_j{\big[1+\lambda+\delta \cos(2\pi{\beta}\times(2j+1))\big]}^2 \bigg\}\nonumber\\
&=\text{sign}\bigg(\prod_j S_1^{2N} - \prod_j S_2^{2N} \bigg),
\end{align}
with
\begin{align}
&\log S_1=\lim\limits_{N\to\infty} \frac{1}{N} \sum_{j=0}^{N} \log|1-\lambda+\delta \cos(2\pi{\beta}\times 2j)|,\\
&\log S_2=\lim\limits_{N\to\infty} \frac{1}{N} \sum_{j=0}^{N} \log|1+\lambda+\delta \cos[2\pi{\beta}\times(2j+1)]|.
\end{align}
Here we set the nearest-neighbor hopping rate $t$ as the unit of energy. For the convenience of discussion, we define
\begin{align}
W=\log S_2-\log S_1.
\end{align}

Applying the Weyl's equidistributional theorem and converting the summations above to integrals, we have the following results
\begin{align}\label{eq:eqnu}
W =
\begin{cases}
>0, & \text{for } \delta\leq 1+\lambda\\
0, & \text{for } \delta > 1+\lambda
\end{cases},
\end{align}
which leads to
\begin{align}\label{eq:eqnu}
\nu =
\begin{cases}
1, & \text{for } \delta\leq 1+\lambda\\
0.5, & \text{for } \delta > 1+\lambda
\end{cases}.
\end{align}
This is consistent with numerical results in Fig.~1(d) of the main text, where both LTM and A-MCD appear to distribute around $\nu=0.5$ in region IV. However, while topological edge states disappear in region IV, the half-integer value of $\nu$ therein does not correspond to the number of edge states under an open boundary condition. We attribute such a reality to the closing of the bulk gap in region IV, which invalidates $\nu$ as a topological invariant.

The analysis above can also be understood in terms of the inverse localization length $\Lambda^{-1}$ calculated from the transfer matrix. More specifically, we consider the Schr\"odinger's equation for the zero-energy state $|\Psi\rangle=\sum_j \Psi(j) c^\dag_j|\text{vac}\rangle$
\begin{align}
{\hat{H}}_{0}|\Psi\rangle= 0,
\end{align}
which leads to
\begin{align}
\Psi(N) = (-1)^{N} \prod_{j=0}^{N-1}\frac{1+\lambda+\delta \cos[2\pi{\beta}\times(2j+1)]}{1-\lambda+\delta \cos(2\pi{\beta}\times 2j)}\Psi(1).
\end{align}

It follows that~\cite{Hughes2014}
\begin{align}
\Lambda^{-1}&=\lim\limits_{N\to\infty} \frac{1}{N}\log|\frac{\Psi(N)}{\Psi(1)}|\\
&=|\lim\limits_{N\to\infty} \frac{1}{N} \sum_{j=0}^{N-1}(\log|1+\lambda+\delta \cos[2\pi{\beta}\times(2j+1)]|-\log|1-\lambda+\delta \cos(2\pi{\beta}\times 2j)|)|\\
&=\left|\log S_2-\log S_1\right|\\
&=\left|W\right|.
\end{align}
Therefore, $\Lambda^{-1}=0$ for $\delta>1+\lambda$, i.e., the localization length diverges in the entire region IV. Physically, such a result derives from the fact that the quasi-periodic disorder is imposed on both the inter- and intra-cell couplings. If, for instance, one adopts a model where quasi-periodic disorder is only imposed on the inter- or intra-cell hopping rates ~\cite{arXiv.2008.06493}, the band gap would open up in the topologically trivial region, with a vanishing topological invariant $\nu$. The localization length then would only diverge on the boundary of the topological phase transition.

\section{ Interaction effect}
\begin{figure}[hbp]
\includegraphics[width= 0.47\textwidth]{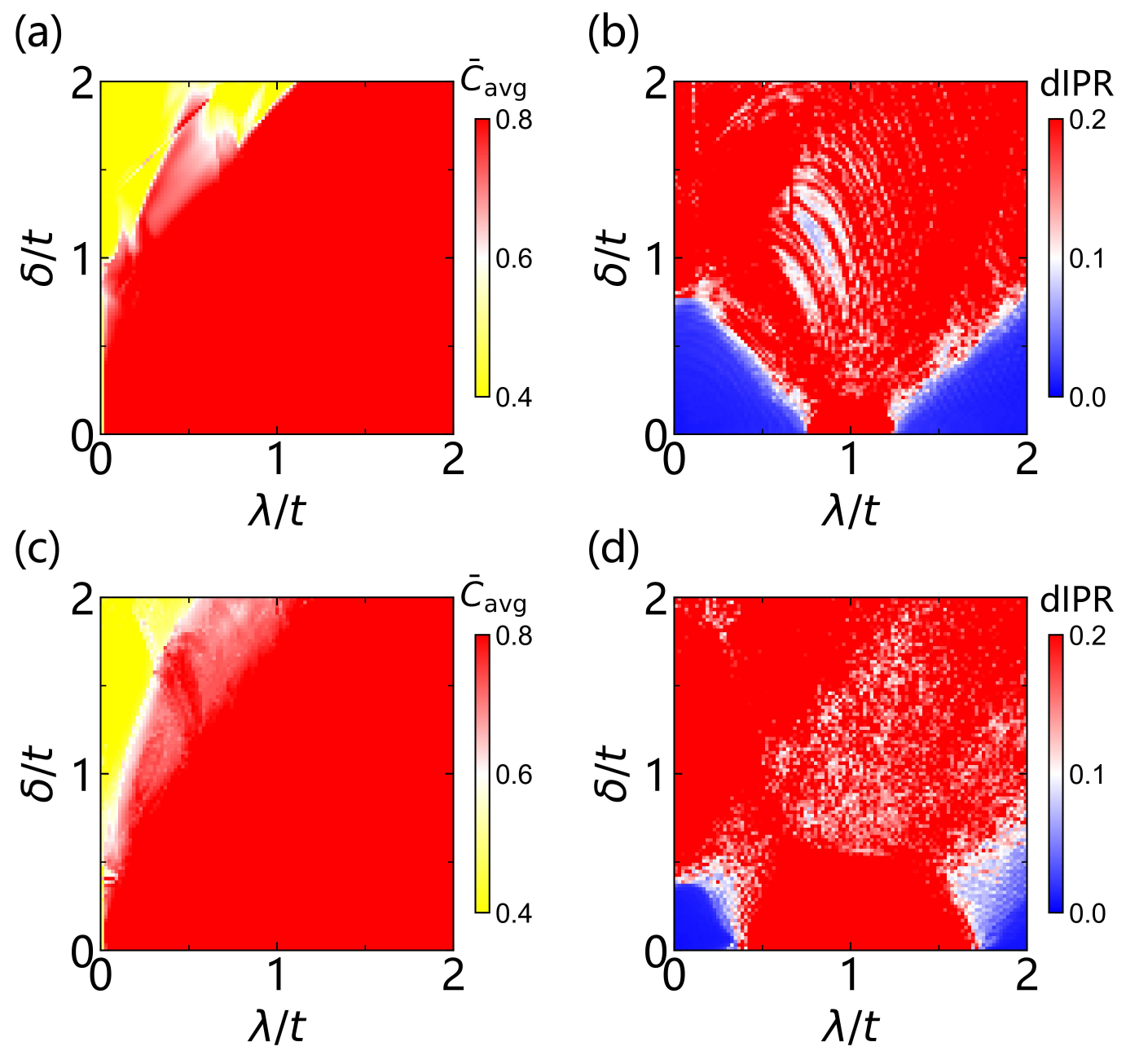}
\caption{\label{fig:S5} Numerical simulation of (a)(c) $\bar{C}_{\text{avg}}$ and (b)(d) $\text{dIPR}$. The interaction parameter is $U/t=0.7$ in (a)(b), and $U/t=2$ in (c)(d). Other parameters are same as those in Fig.~\ref{fig:S4}.
}
\end{figure}
Eq.~(6) in the main text provides a qualitatively valid description of the interaction effects, under the assumptions: i) all momentum states occupy the same spatial mode; ii) different momentum states can be considered as distinguishable quantum states; iii) quantum fluctuations are negligible. For our experiment, conditions i) and ii) are met so long as momentum lattice sites can be considered as distinguishable quantum states. This is guaranteed by the decompression process. The mean-field condition is also satisfied since the average number density per site is much larger than unity. Indeed, as demonstrated by Fig.~5(c) of the main text,
the numerically calculated A-MCD [using Eq.~(6)] qualitatively matches the measured A-MCD under different interactions.

To supplement Fig.~5 of the main text in depicting the interaction effects, we numerically simulate dynamics under the mean-field Hamiltonian $\hat{H_0}+\hat{H}_{\text{int}}$ for a longer evolution time of $60 \hbar/t$, with either small [Fig.~\ref{fig:S5}(a)(b)] or large [Fig.~\ref{fig:S5}(c)(d)] interaction energy. The resulting topological and localization phase diagrams in terms of $\bar{C}_{\text{avg}}$ and dIPR are shown in Fig.~\ref{fig:S5}(a)(c) and (b)(d), respectively.
Compared with the non-interacting case in Fig.~\ref{fig:S4}, phase diagrams in Fig.~\ref{fig:S5}(a)(b) are almost the same, indicating the robustness of phase boundaries under small interactions.
However, both the topological and localization phase boundaries are shifted under larger interactions, leading to a larger stability region of the critically localized topological phase in Fig.~\ref{fig:S5}(c)(d). The long-time numerical simulation thus supports our conclusions in the main text.


\bibliography{GAAHdraft_1022}